\documentclass{PoS}

\title{Heavy baryon mass spectrum from lattice QCD with 2+1 dynamical sea quark flavors}

\ShortTitle{Heavy baryon mass spectrum}

\author{\speaker{Heechang Na} and Steven Gottlieb\\
        Department of Physics, Indiana University, Bloomington, Indiana 47405, USA\\
        E-mail: \email{heena@indiana.edu}, \email{sg@indiana.edu}}

\abstract{We investigate the mass differences of charm and bottom heavy baryons, using MILC lattice gauge 
configurations with 2+1 dynamical sea quark flavors. We extend our previous work to three lattice 
spacings: fine ($a \approx 0.09$), coarse ($a \approx 0.12$), and medium-coarse ($a \approx 0.15$) ensembles. For extrapolations 
and interpolations, we apply simultaneous quadratic fits and simple linear fits with the full QCD data points for the valence quark masses 
as well as the sea quark masses.}

\FullConference{The XXVI International Symposium on Lattice Field Theory\\
		 July 14-19 2008\\
		 Williamsburg, Virginia, USA}

\begin{document}

\section{Introduction}
Heavy baryons have been investigated by both experimental and theoretical approaches.
From experiment, the singly charmed heavy baryon 
mass spectrum is well known; however, the other heavy baryon masses are 
only crudely known.  
Recently,  mass measurements of the singly bottom baryon $\Xi_b^-$ have been published by the D$\emptyset$~\cite{D0}  and CDF~\cite{CDF} collaborations. 
From lattice QCD, there are several quenched 
calculations [3--8] and dynamical sea quark flavor simulations [9--11] for the heavy baryon mass spectrum,
and most results are in fair agreement with observed values.
In this work, we extend our previous dynamical sea quark flavor simulation~\cite{na1}\cite{na2} with a larger data set and more systematic analysis methods.
 
 We use the MILC fine ($a \approx 0.09$), coarse ($a \approx 0.12$), and medium-coarse ($a \approx 0.15$) lattices~\cite{lat_parm}.
We apply two local interpolating operators~\cite{ukqcd} and construct heavy baryon two point functions with improved clover heavy quark propagators with the Fermilab interpretation~\cite{ekm} and improved staggered light quark propagators. 
We extract the mass differences from the ratio of propagators and extrapolate to the chiral limit using simultaneous quadratic fits and simple linear fits with the full QCD data points.

\section{Formalism}
We used two types of interpolating operators~\cite{ukqcd} for the singly heavy baryons, which are
 \begin{equation}
\mathcal{O}_5 = \epsilon_{abc} ( \psi_1^{aT} C \gamma_5 \psi_2^b ) \Psi_H^c  \;\;\;\;\textrm{and}\;\;\;
\mathcal{O}_\mu = \epsilon_{abc} ( \psi_1^{aT} C \gamma_\mu \psi_2^b ) \Psi_H^c,
\label{operator}
\end{equation}
where $\epsilon_{abc}$ is the Levi-Civita tensor, $\psi_1$
and $\psi_2$ are light valence quark fields for up, down, or strange quarks,
$\Psi_H$ is the heavy valence quark field for the charm or the bottom quark,
 $C$ is the charge conjugation matrix, and $a$, $b$, and $c$ are color indices.
Basically, $\mathcal{O}_5$ is the operator for
$s^\pi = 0^+$ and  $\mathcal{O}_\mu$ is for
$s^\pi = 1^+$, where $s^\pi$ is the spin parity state of the light quark pair.
Therefore, $\mathcal{O}_5$ describes total spin $J=\frac{1}{2}$ baryons ($\Lambda_H$, $\Xi_H$), and $\mathcal{O}_\mu$ describes $J=\frac{1}{2}$ baryons ($\Sigma_H$, $\Xi^\prime_H$, $\Omega_H$) as well as $J=\frac{3}{2}$ baryons ($\Sigma_H^\ast$, $\Xi^{\prime\ast}_H$, $\Omega_H^\ast$).
For doubly heavy baryons, we can simply interchange the heavy quark field and the light quark fields in Eq.~\ref{operator}~\cite{woloshyn2}.

We apply the method of Wingate \emph{et al}.~\cite{wingate} to combine staggered propagators for the light valence quarks and a Wilson type (clover) propagator for the heavy valence quark.
Since we use staggered light quarks, we needed to consider taste mixing.
In our previous work~\cite{na2}\cite{nagata}, we found that there is no taste mixing between the light quarks. 
However, due to cancellations among the copy indices, we cannot separate spin $J=\frac{1}{2}$ and $J=\frac{3}{2}$ states from the operator $\mathcal{O}_\mu$ using standard spin projection operators.

\section{Data analysis}
Since we use the Fermilab interpretation~\cite{ekm} for the heavy quarks, the absolute mass from the simulation is not a physical quantity. The physical mass $M_{phy}$ of the baryon is
\begin{equation}
\label{ }
M_{phy}=M_{cal} + \Delta,
\end{equation}
where $M_{cal}$ is the simulation result and $\Delta$ is a constant mass shift.
We can calculate the constant mass shift from calculations of the kinetic mass with non-zero momenta or heavy-light meson spectroscopy. 
In this project, however, we estimate mass differences, which are directly measurable quantities without determining $\Delta$. 

Since we are interested in mass differences, we can take the ratio of two propagators and fit the mass difference directly.
A fit model function $P_r(t)$ of the ratio of the propagators is\begin{equation}
\label{ }
P_r(t) = \frac{A_1 e^{-m_1 t} + \cdots}{ A_2 e^{-m_2 t} + \cdots } = A_1^\prime e^{- (m_1 - m_2) t} + \cdots.
\end{equation}
Since the propagators are calculated using the same gauge configurations, the propagators in numerator and denominator are correlated with each other. 
Using this method, we expect smaller statistical errors and more stable fits.

We can take the ratio of the propagators, for example, between $\Xi_c^\prime$ and $\Lambda_c$,
\begin{equation}
\label{ }
C_r(t) = \frac{\sum_{k=1}^N C_k^{\Xi_c^\prime}(t)}{ \sum_{k=1}^N C_k^{\Lambda_c}(t) } = \frac{\bar{C}^{\Xi_c^\prime}(t)}{\bar{C}^{\Lambda_c}(t)},
\end{equation}
where $k$ is the configuration index.
The covariance matrix of the ratio is estimated using bootstrap sampling.

\begin{figure}[htp]
\centering
\includegraphics[width=.47\textwidth]{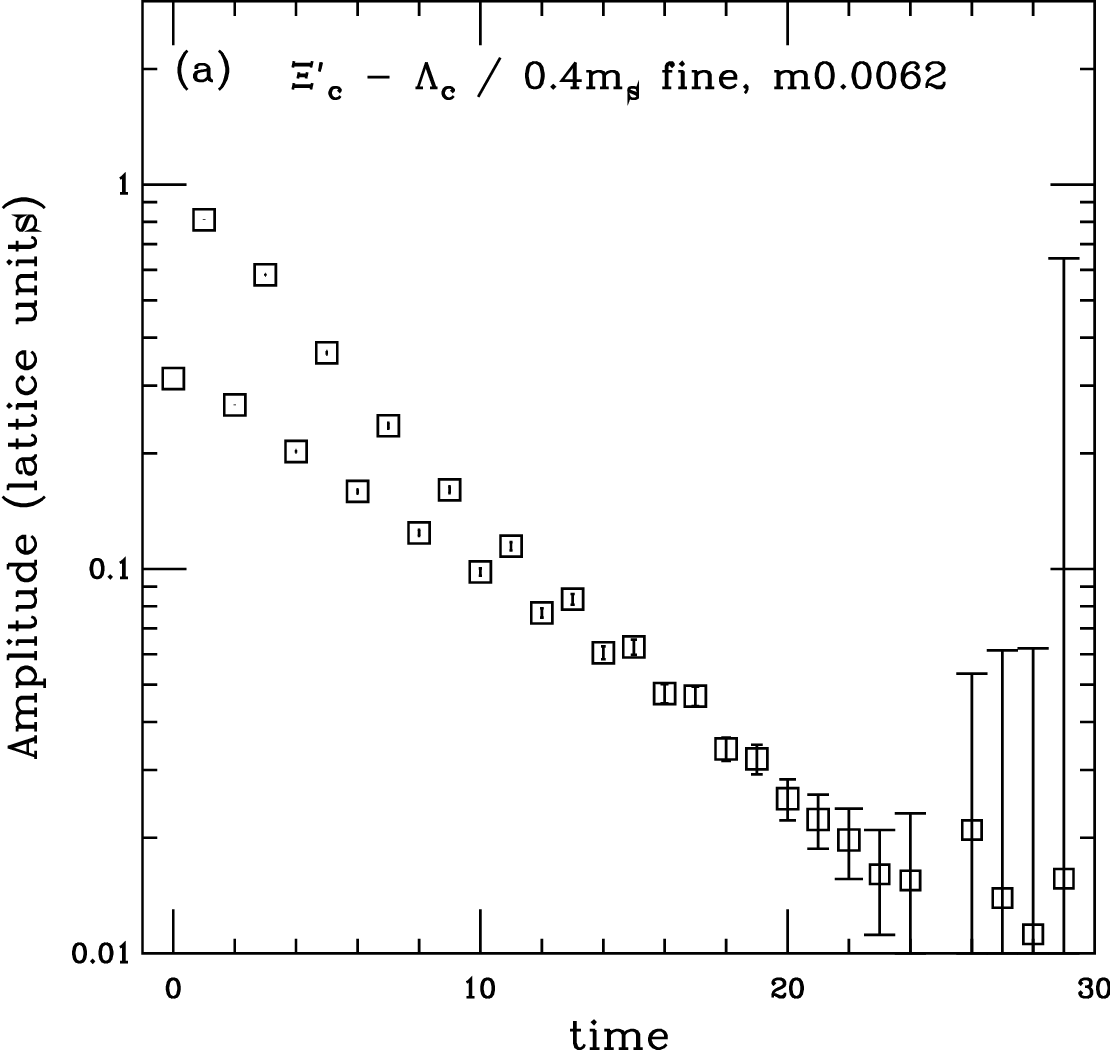}
\includegraphics[width=.46\textwidth]{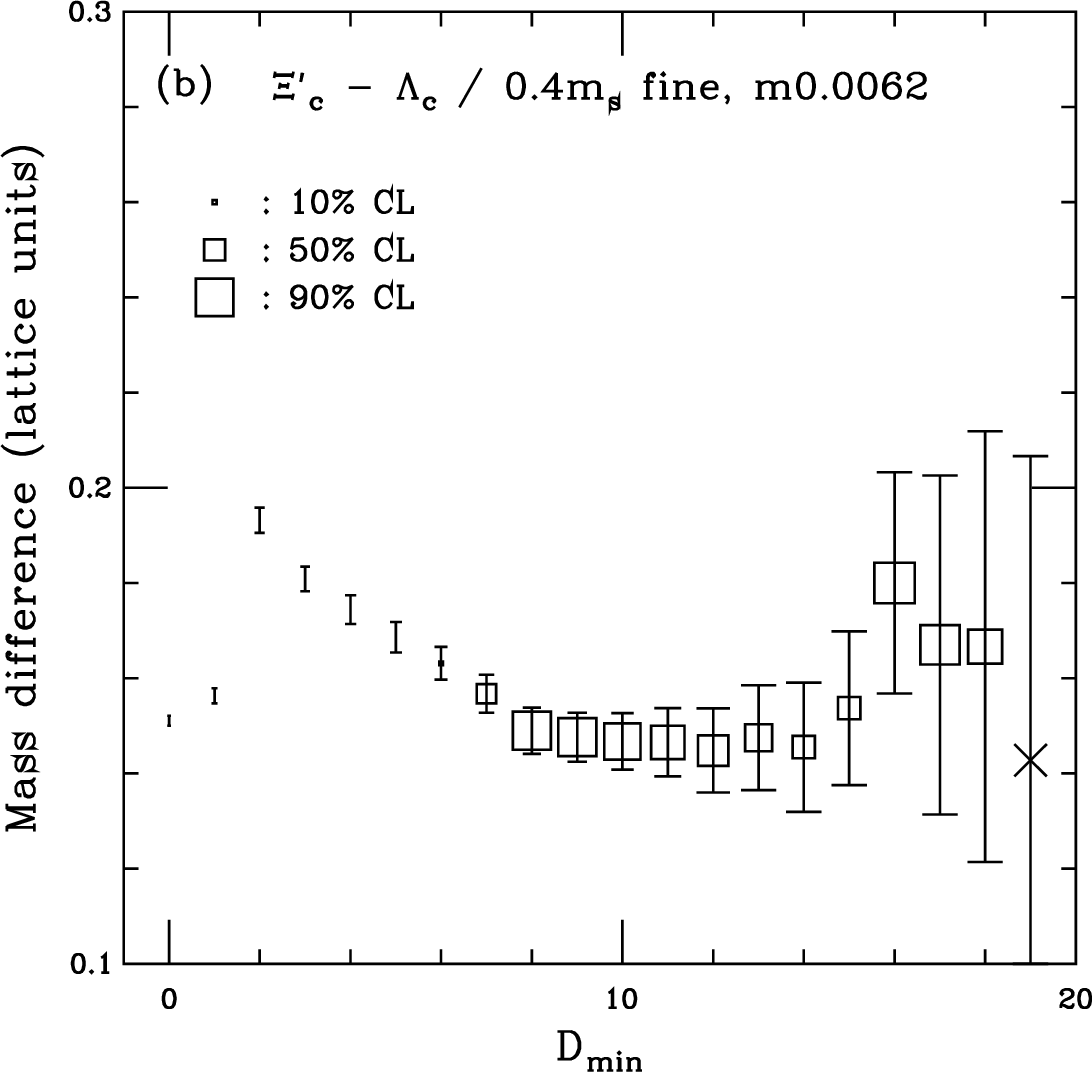}
\caption{The ratio of the propagators (a) and the two particle fit of the ratio (b). The size of the symbols on (b) is proportional to the confidence level (CL).}
\label{fig:prop_ratio}
\end{figure}

We display an example of this method for $\Xi_c^\prime$ and $\Lambda_c$ on the fine lattice with dynamical quark masses $m_l = 0.4 m_s$ and 0.0062 light valence quark mass in Fig.~\ref{fig:prop_ratio}.
The jittery pattern appears strongly in the ratio (Fig.~\ref{fig:prop_ratio} (a)), because the opposite parity state contributions are substantially different for numerator and denominator.
In fact, $\Xi_c^\prime$ contains large opposite parity state contributions, while $\Lambda_c$ contains very weak opposite parity state contributions.

We obtained the most reasonable fit from a fit model function with one non-alternating phase exponential and one alternating phase exponential.
That fit is shown in Fig.~\ref{fig:prop_ratio}(b).
The maximum distance of the fit is $D_{\rm{max}}=22$ and a plateau appears where $D_{\rm{min}}\geq 8$.

\begin{figure}[htp]
\centering
\includegraphics[width=.47\textwidth]{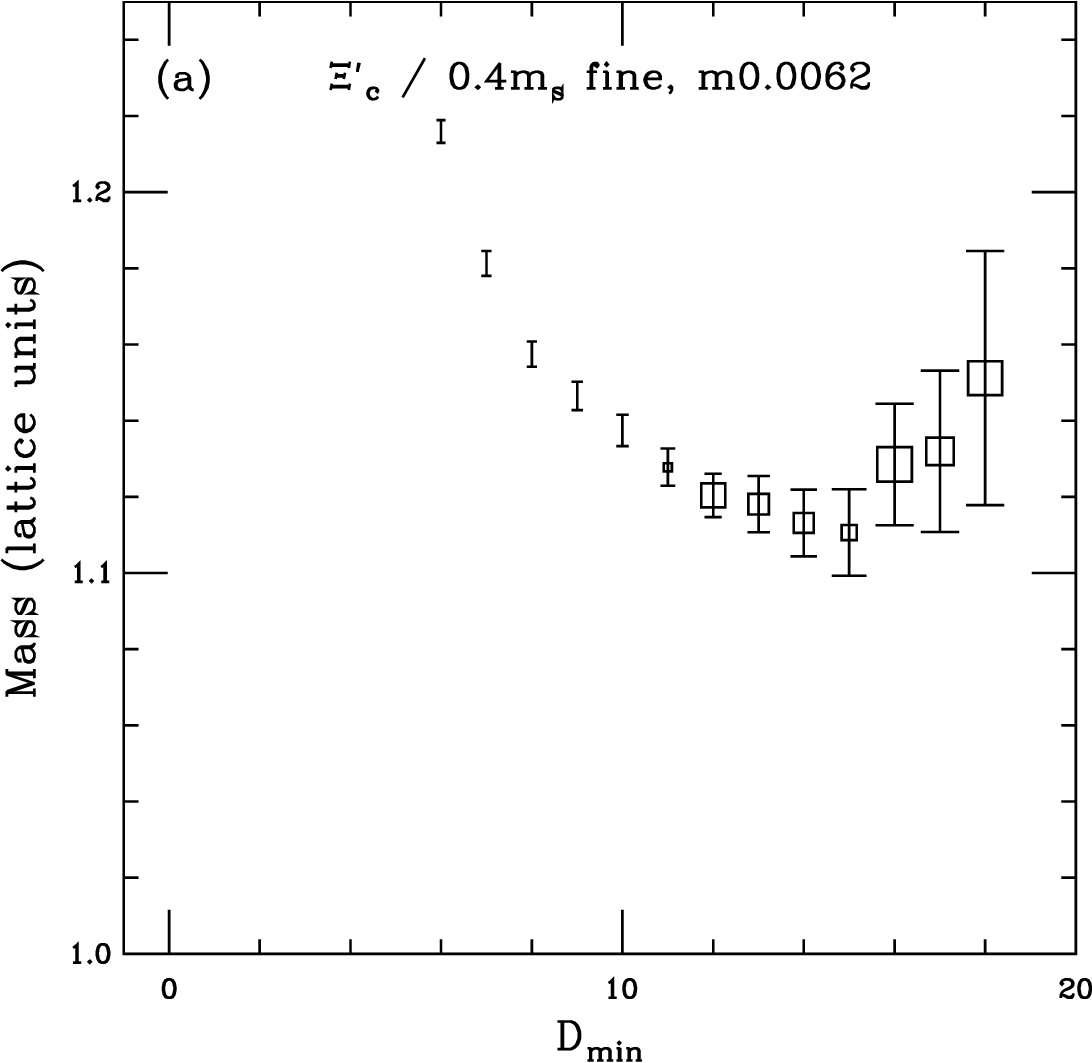}
\includegraphics[width=.47\textwidth]{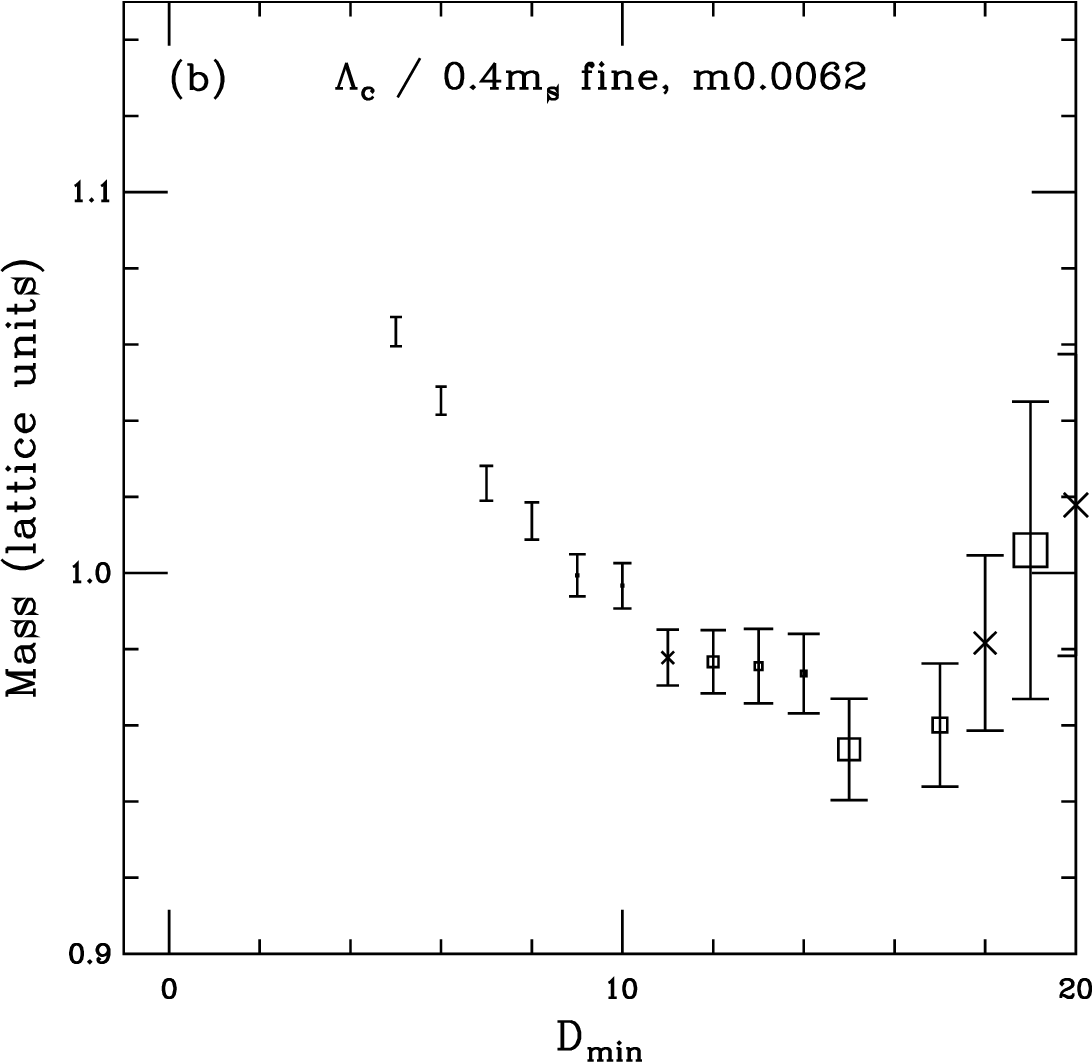}
\caption{The individual fits for $\Xi_c^\prime$ (a) and $\Lambda_c$ from the propagators. The square symbols indicate successful fits and the cross symbols indicate unstable fits. The size of the symbols is proportional to the confidence level. We obtained these fits from a fit model function with one non-alternating phase exponential and one alternating phase exponential.}
\label{fig:individual_mass}
\end{figure}

The mass difference can also be obtained by first fitting the propagators individually, and then subtracting the masses to obtain the difference.
We can compare this method shown in Fig.~\ref{fig:individual_mass} to the ratio method.
As we expected, the individual mass fits for $\Xi_c^\prime$ and $\Lambda_c$ (Fig.~\ref{fig:individual_mass}) are worse than the fit of the ratio (Fig.~\ref{fig:prop_ratio} (b)).
For $\Xi_c^\prime$, the confidence levels are fine; however, the fluctuation of the plateau is larger (the systematic error is larger), and for $\Lambda_c$, fits are not stable.
From this comparison, we conclude that the ratio method is better.

Once we obtain the mass differences, we need to perform extrapolations to the physical quark mass.
In this analysis, we extrapolate the valence quark masses and the sea quark masses simultaneously, using a quadratic fit model function given by
\begin{equation}
\label{quad_fit}P_{\rm{quad}} = c_0 + c_1 m_l + c_2 m_l^2 + c_3 m_s + c_4 m_{\rm{sea}}
\end{equation}where $c_0$ to $c_4$ are the fitting parameters, $m_l$ is the light valence quark mass, $m_s$ is the strange valence quark mass, and $m_{\rm{sea}}$ is the light sea quark mass.
We fixed $c_3=0$ except for the coarse lattice, since only the coarse lattice has multiple (three) strange valence quark masses.
We would prefer to use a heavy baryon mass formula based on partially quenched chiral perturbation theory, but do not know of one.
We can also perform the extrapolations with only the full QCD data points, \emph{i.e.}, the points for which the valence quark masses match the corresponding sea quark masses.
In general, the errors of the full QCD extrapolations are larger than those of the simultaneous quadratic fits, because the simultaneous fit uses more data to constrain the fit parameters.

\section{Results and discussion}
Parameters of the ensembles~\cite{lat_parm} are summarized in Table~\ref{tab:param}.
The scale of each ensemble was determined by a length scale $r_1$ from the static quark potential using $f_\pi$ as an input parameter~\cite{r1}:
\begin{equation}
r_1 = 0.3108 (15) (^{+26}_{-79}) \rm{fm}. 
\end{equation}
\begin{table}[tp] 
\centering
\begin{tabular}{|c c c c c c c|}
\hline
a (\rm{fm})&$am_l/am_s$&L (\rm{fm})&$\beta$&size&$r_1/a$&\# confs\\
\hline
$\sim$0.15 & 0.0097 / 0.0484 & 
2.4 & 6.572 &
$16^3\times 48$ & 2.1356&631\\
$\sim$0.15 & 0.0194 / 0.0484 &
2.4 & 6.586 &
$16^3\times 48$ & 2.1284&631\\$\sim$0.15 & 0.0290 / 0.0484 &
2.4 & 6.600 &$16^3\times 48$ & 2.1245&440\\
\hline
$\sim$0.12 & 0.007 / 0.05 &
2.4 & 6.76 &
$20^3\times 64$ & 2.6349&545\\
$\sim$0.12 & 0.01 / 0.05 &
2.4 & 6.76 &
$20^3\times 64$ & 2.6183&590\\
$\sim$0.12 & 0.02 / 0.05 &
2.4 & 6.79 &
$20^3\times 64$ & 2.6477&452\\
\hline
$\sim$0.09 & 0.0062 / 0.031 &
2.4 & 7.09 &
$28^3\times 96$ & 3.7016&534\\
$\sim$0.09 & 0.0124 / 0.031 &
2.4 & 7.11 &
$28^3\times 96$ & 3.7138&557\\
\hline
\end{tabular}
\caption{MILC lattice parameters.  The three nominal lattice
spacings in the first column (0.15, 0.12, and 0.09 fm) correspond to
medium-coarse, coarse, and fine ensembles, respectively.  The bare light
(strange) sea quark mass in lattice units is $am_l$ ($am_s$).  The
spatial size of the lattice is $L$ and $\beta=10/g^2$, where $g$ is the
bare gauge coupling.  In the sixth column we show $r_1/a$ calculated
from a global fit to $r_1$ values on all our ensembles~\cite{doug}.}
\label{tab:param}
\end{table}

\begin{figure}[htp]
\centering
\includegraphics[width=.47\textwidth]{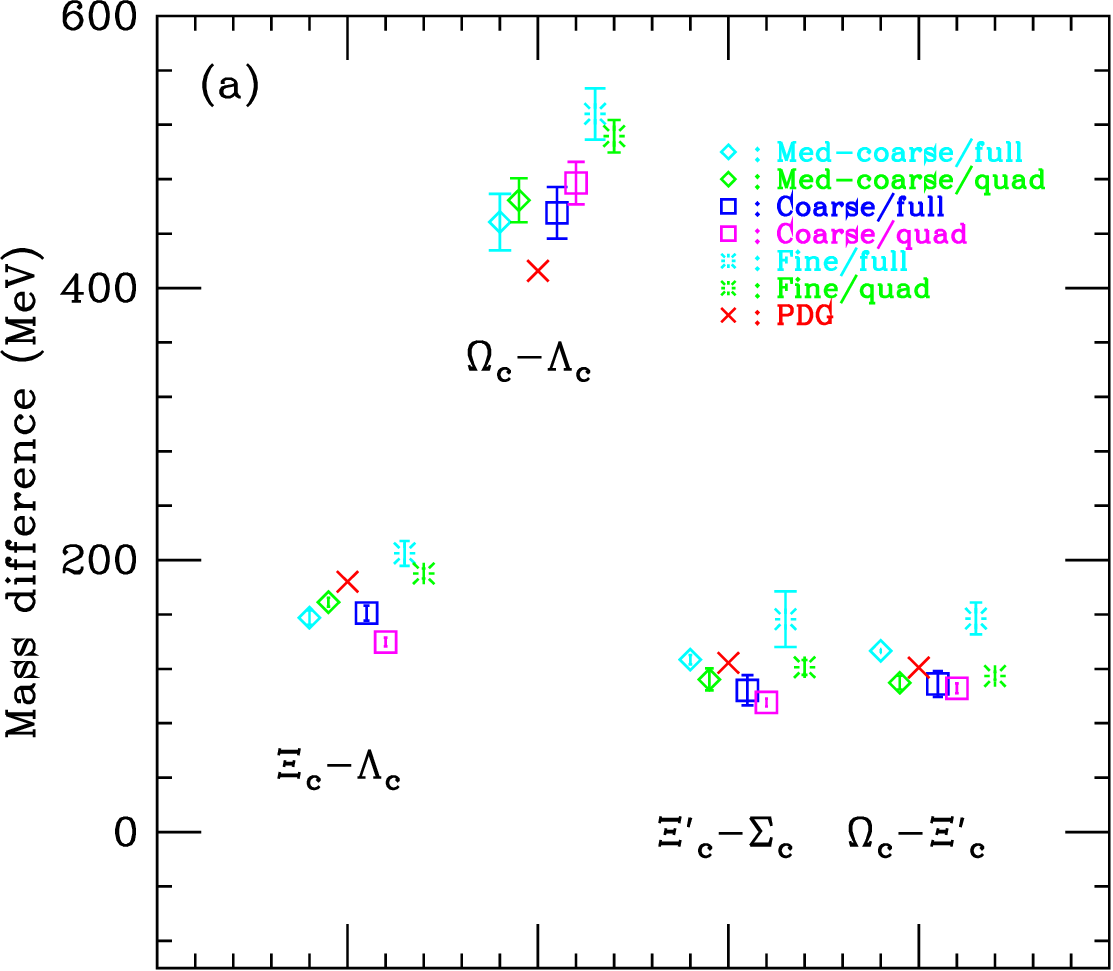}
\includegraphics[width=.47\textwidth]{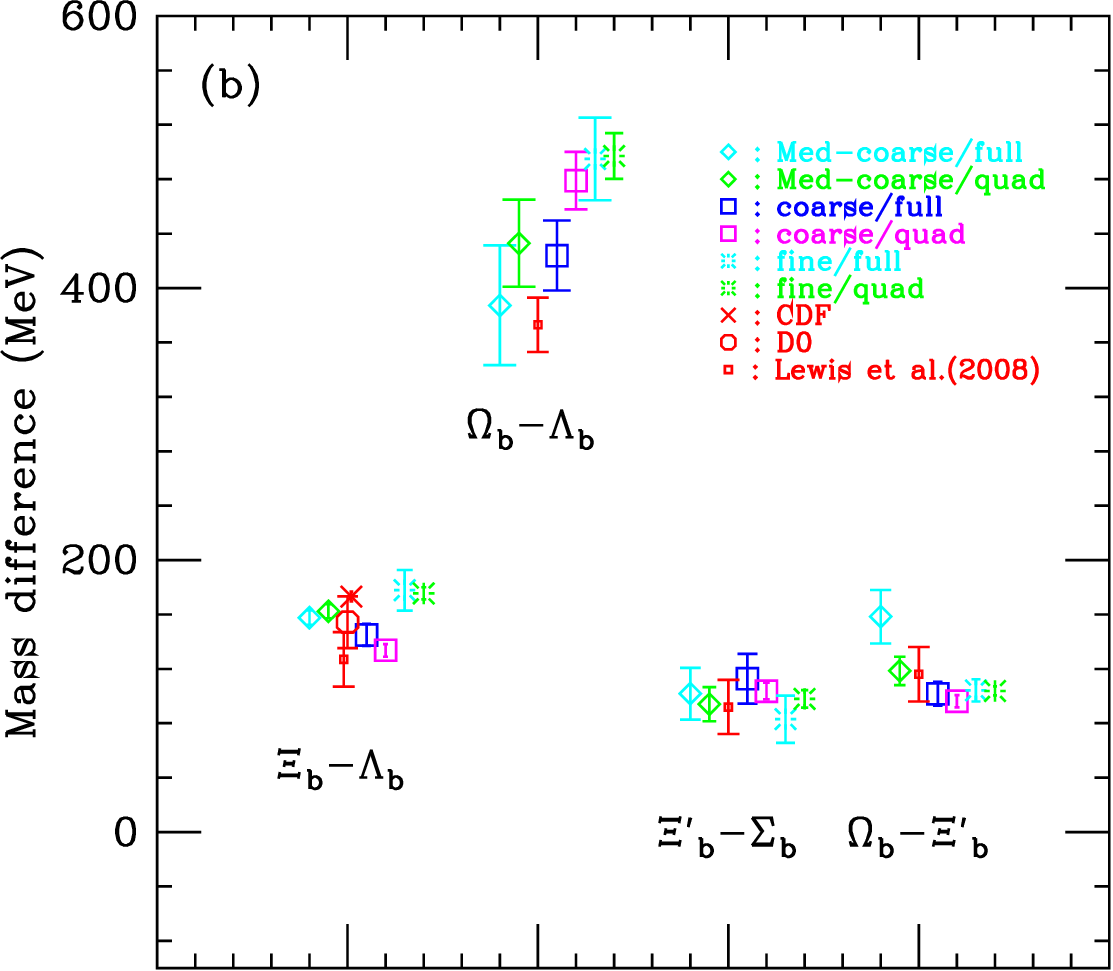}
\caption{Independent mass differences of $J^p=\frac{1}{2}^+$ singly charmed baryons (a), and singly bottom baryons (b). In the legend, ``full'' indicates the full QCD data point fit result, and ``quad'' indicates the simultaneous quadratic fit result.  }
\label{fig:singly}
\end{figure}

We present mass splittings in MeV for singly heavy baryons in Fig.~\ref{fig:singly}.
We investigate five $J^p=\frac{1}{2}^+$ singly heavy baryons ($\Lambda_H$, $\Xi_H$, $\Sigma_H$, $\Xi^\prime_H$, $\Omega_H$).
There are ten possible mass splitting combinations.
Since not all of the combinations are independent and all dependent combinations are consistent each other, we display only four independent combinations on the figure. 
We display the comparison of the singly charmed baryons and the PDG data~\cite{pdg} in Fig.~\ref{fig:singly} (a).
The error bars of our calculations are larger than those of experiment, but smaller than or comparable to those of previous lattice calculations~[3--8].
For the singly bottom baryons in Fig.~\ref{fig:singly} (b), we compare with experimental results measured by the CDF and D$\emptyset$ collaborations~\cite{D0}\cite{CDF} and a dynamical calculation result by Lewis and Woloshyn~\cite{woloshyn_dynamic}. 
Our error bars, in general, are smaller than those of Lewis and Woloshyn, especially for the fine lattice result.

Since the singly charmed baryons are experimentally well known, let us examine more closely our singly charm results.
The mass differences between $\Xi_c-\Lambda_c$, $\Xi_c^\prime - \Sigma_c$, and $\Omega_c - \Xi_c^\prime$ are in good agreement with the PDG data~\cite{pdg}; however, the mass difference  $\Omega_c - \Lambda_c$ is not.
Note that $\Lambda_c$ and $\Xi_c$ are calculated from the operator $\mathcal{O}_5$, and  $\Sigma_c$, $\Xi_c^\prime$, and $\Omega_c$ are calculated from the operator $\mathcal{O}_\mu$.
Thus, the disagreement occurs on the mass differences of the heavy baryons that come from the different operators.
On the fine lattice result, especially, this pattern is quite evident.
We checked the other mass differences with the different operator combinations, and we obtained similar discrepancies from the PDG data~\cite{pdg}.
We hope that this can be resolved by studying the constant mass shift due to the heavy quarks or the hyper-fine structure of the singly heavy baryons.
Resolving this puzzle is a high priority.

\begin{figure}[htp]
\centering
\includegraphics[width=.47\textwidth]{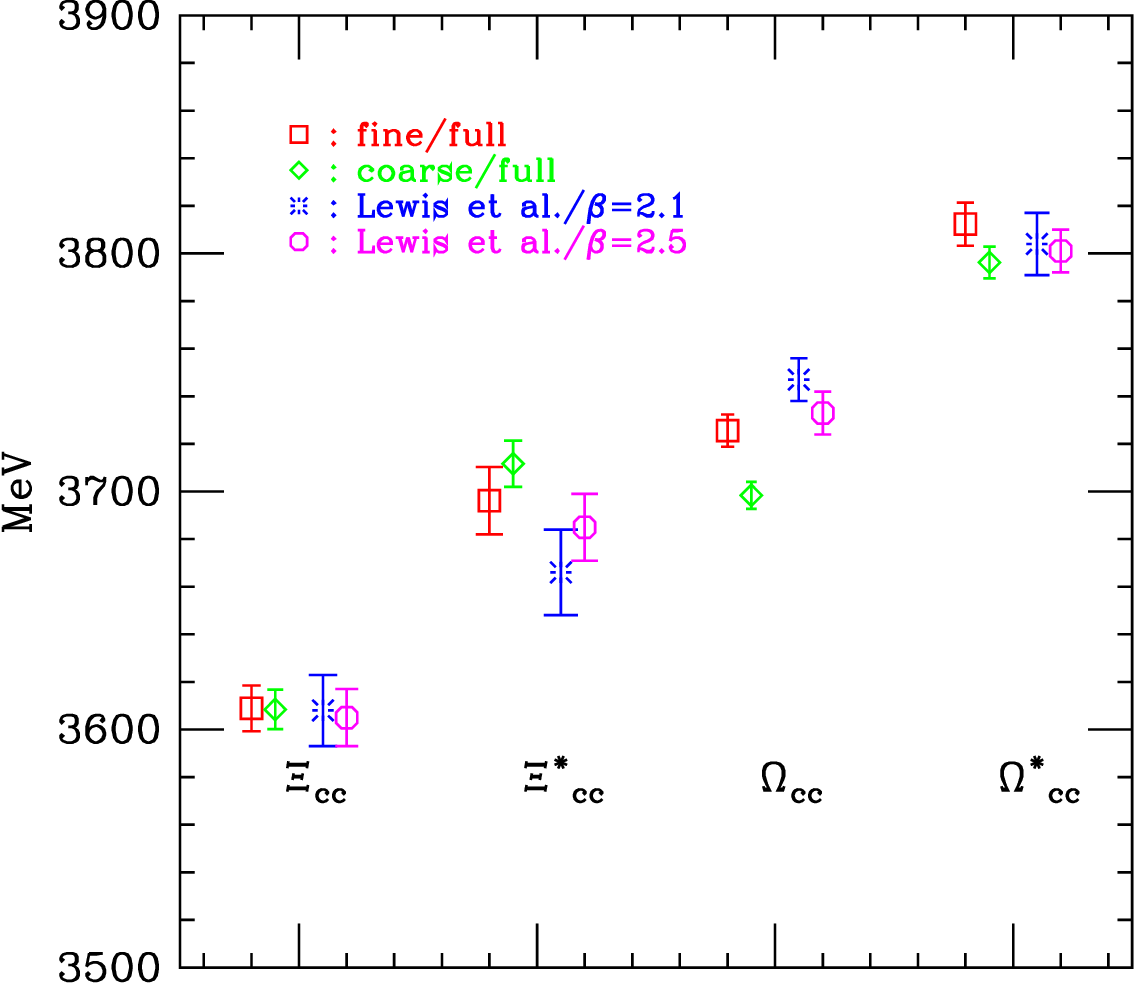}
\includegraphics[width=.47\textwidth]{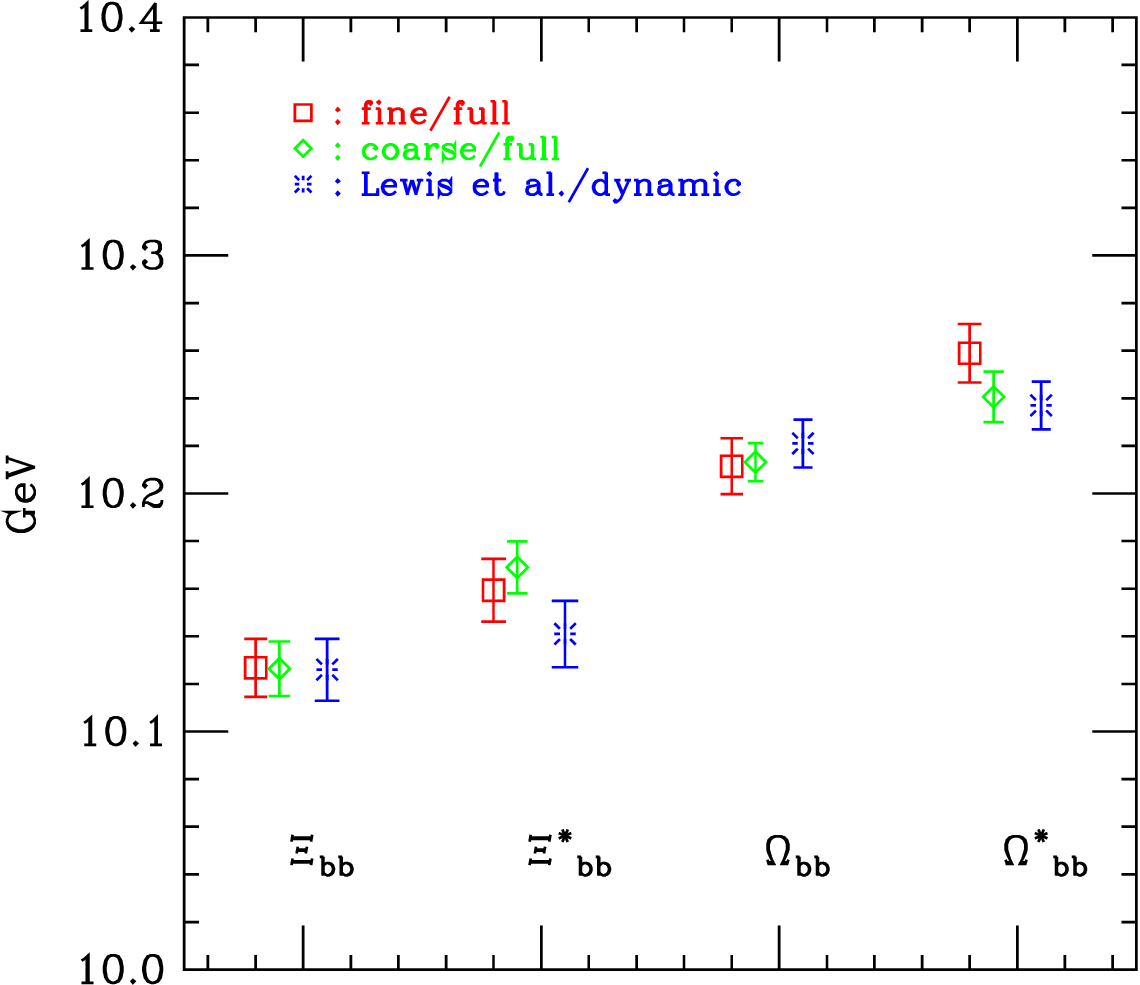}
\caption{The mass spectrum of doubly charmed and bottom baryons. The error bars are statistical only.}
\label{fig:doubly_heavy}
\end{figure}

We also present the doubly charmed and bottom baryons in Fig.~\ref{fig:doubly_heavy}.
We display our fine and coarse lattice results with those of Lewis \emph{et al}.~\cite{woloshyn1} for the doubly charmed baryons and Lewis and Woloshyn~\cite{woloshyn_dynamic} with dynamical sea quarks for the doubly bottom baryons.
We set the scale of the $y$-axis of the figure from the other group's results, because they calculated the absolute masses, while we calculated the static masses.
For the doubly charmed baryons, we added 193 and 438 MeV, and for the doubly bottom baryons, we added 2.833 and 4.223 GeV to our fine and coarse lattice results, respectively.
Our results agree fairly well with prior results.
The mass differences between $\Xi_{HH}$ and $\Omega_{HH}$ are about 100 MeV, which is about the strange quark mass.
The hyper-fine splittings are about $30 \sim 80$ MeV.
Moreover, the hyper-fine splittings of the doubly bottom baryons are smaller than those of the doubly charmed baryons, because of heavy quark symmetry.

\newpage

\acknowledgments{Numerical calculations were performed on the Kaon and Pion clusters at Fermilab and on the BigRed cluster at Indiana University. This work was supported in part by the U.S. DOE under grant DE-FG02-91ER--40661.}

\end{document}